\documentclass[twocolumn,showpacs,preprintnumbers,amsmath,amssymb]{revtex4-1}
\usepackage{graphicx}
\usepackage{rotating}

\begin{document}
\renewcommand{\theequation}{\thesection.\arabic{equation}}
\newcommand{\re}{\mathop{\mathrm{Re}}}
\newcommand{\be}{\begin{equation}}
\newcommand{\ee}{\end{equation}}
\newcommand{\bea}{\begin{eqnarray}}
\newcommand{\eea}{\end{eqnarray}}
\title{Redshift drift test of exotic singularity universes}
\author{Tomasz Denkiewicz}
\email{atomekd@wmf.univ.szczecin.pl}
\affiliation{\it Institute of Physics, University of Szczecin, Wielkopolska 15, 70-451 Szczecin, Poland}
\affiliation{\it Copernicus Center for Interdisciplinary Studies, S{\l }awkowska 17, 31-016 Krak\'ow, Poland}
\author{Mariusz P. D\c{a}browski}
\email{mpdabfz@wmf.univ.szczecin.pl}
\affiliation{\it Institute of Physics, University of Szczecin, Wielkopolska 15, 70-451 Szczecin, Poland}
\affiliation{\it Copernicus Center for Interdisciplinary Studies, S{\l }awkowska 17, 31-016 Krak\'ow, Poland}
\author{Carlos J. A. P. Martins}
\email{Carlos.Martins@astro.up.pt}
\affiliation{\it Centro de Astrofisica da Universidade do Porto, Rua das Estrelas, 4150-762 Porto, Portugal}
\author{Pauline E. Vielzeuf}
\email{Pauline.Vielzeuf@astro.up.pt}
\affiliation{\it Centro de Astrofisica da Universidade do Porto, Rua das Estrelas, 4150-762 Porto, Portugal}

\date{\today}
\input epsf
\begin{abstract}

We discuss how dynamical dark energy universes with exotic singularities may be distinguished from the standard $\Lambda$CDM model on the basis of their redshift drift signal, for which measurements both in the acceleration phase and in the deep matter era will be provided by forthcoming astrophysical facilities. Two specific classes of exotic singularity models are studied: sudden future singularity models and finite scale factor singularity models. In each class we identify the models which can mimic $\Lambda$CDM and play the role of dark energy as well as models for which redshift drift signals are significantly different from $\Lambda$CDM and the test can differentiate between them.

\end{abstract}


\maketitle

\section{Introduction}
\setcounter{equation}{0}

Exotic singularities are all the non-big-bang/-big-crunch types of singularities which started being investigated after the discovery of the accelerated expansion of the universe \cite{supernovaold}. Their classification was first attempted in Ref. \cite{nojiri} and further extended in Refs. \cite{AIP2010,sesto}. The list is pretty long and includes a big-rip \cite{phantom}, a little rip \cite{LRip}, a pseudo-rip \cite{PRip}, a sudden future singularity \cite{BGT,Varun,barrow04}, a generalized sudden future singularity \cite{GSFS}, a finite scale factor singularity \cite{nojiri,JCAP12,ABTD}, a big-separation \cite{nojiri}, a $w$-singularity \cite{wsing}, and their generalizations \cite{yurov}. All of them but big-rip and little-rip are geodesically complete and do not fulfil the standard requirement for being a singularity required by the Hawking-Penrose theorems. The geodesic equations do not feel these singularities \cite{LFJ} and due to this, even extended objects may pass
  through them without being destroyed \cite{adam}. This property allows some of exotic singularities to show up in the near future with no contradiction to observational data and in such a way they may serve as a kind of dark energy which leads to the acceleration of the universe.

In this paper two particular examples of exotic singularities will be studied. The first is a sudden future singularity (SFS, which seemingly appear uniquely in loop quantum cosmology \cite{LQC}) which has been shown to be observationally allowed \cite{DHD,GHDD,DDGH,laszlo}, and the second is a finite scale factor singularity (FSFS) which was also checked against standard cosmological data (supernovae, baryon acoustic oscillations, CMB shift parameter) \cite{JCAP12,ABTD}. Some other before mentioned types of exotic singularities have also been tested (e.g. \cite{f(rho)}). Here we are going to concentrate on an astrophysical observable known as the redshift drift \cite{sandage}. This test has been discussed in many other cosmological frameworks \cite{Quercellini12,pauline12}, but not yet in the context of exotic singularities. We want to fill in this gap and discuss it in the present contribution.

The paper is organized as follows. In Section \ref{exotic} we briefly review the basic properties of exotic singularities under study. In Section \ref{drift} we provide a short overview of redshift measurements, and in Sect. \ref{results} we discuss its signatures for these singularities. In Section \ref{rac} we present our conclusions.

\section{SFS and FSFS in isotropic cosmology}
\setcounter{equation}{0}
\label{exotic}

SFS and FSFS singularities show up within the framework of Einstein-Friedmann cosmology governed by the
standard field equations
\bea \label{rho} \varrho(t) &=& \frac{3}{8\pi G}
\left(\frac{\dot{a}^2}{a^2} + \frac{kc^2}{a^2}
\right)~,\\
\label{p} p(t) &=& - \frac{c^2}{8\pi G} \left(2 \frac{\ddot{a}}{a} + \frac{\dot{a}^2}{a^2} + \frac{kc^2}{a^2} \right)~,
\eea
where the energy-momentum conservation law
\be
\label{conser}
\dot{\varrho}(t) = - 3 \frac{\dot{a}}{a}
\left(\varrho(t) + \frac{p(t)}{c^2} \right)~,
\ee
is trivially fulfilled due to the Bianchi identity. Here $a \equiv a(t)$ is the scale factor, the dot means the derivative with respect to time $t$, $G$ is the gravitational constant, $c$ is the velocity of light, and the curvature index $k=0, \pm 1$. For further considerations we set $k=0$, and define an effective barotropic index as \cite{DDGH}
\be
\label{weff}
w_{eff} = \frac{p}{\varrho c^2} = w_{eff}(t) = \frac{1}{3} \left( 2 q(t) -1 \right)~~,
\ee
where $q(t) = - \ddot{a} a/\dot{a}^2$ is the deceleration parameter. In both cases of SFS and FSFS singularities we take the scale factor in the form
\be
\label{sf2} a(t) = a_s \left[\delta + \left(1 - \delta \right) \left( \frac{t}{t_s} \right)^m - \delta \left( 1 - \frac{t}{t_s} \right)^n \right]~,
\ee
with the appropriate choice of the constants $\delta, t_s, a_s, m,n$ \cite{barrow04,DHD}. In order to have accelerated expansion in an SFS universe, $\delta$ has to be negative ($\delta<0$), and to have such an effect in an FSFS universe, $\delta$ has to be positive ($\delta>0$). For $1<n<2$ we have an SFS, while in order to have an FSFS, $n$ has to be in the range $0<n<1$. As can be seen from (\ref{rho})-(\ref{sf2}), for SFS at $t=t_s$, $a \to a_s$, $\varrho \to \varrho_s=$ conts., $p \to \infty$, while for an FSFS the energy density $\rho$ also diverges and we have: for $t\rightarrow t_s$, $a\rightarrow a_s$, $\rho\rightarrow\infty$, and $p \rightarrow \infty$, where $a_s,\ t_s,\ \rho_s$, are constants and $a_s\neq 0$. A special case of an SFS in which the anti-Chaplygin gas equation of state is allowed \cite{kamenshchik}
\be
p(t) = \frac{A}{\varrho(t)} \hspace{0.5cm} (A \geq 0)~~,
\label{eoschap}
\ee
is a big-brake singularity for which $\varrho \to 0$ and $p \to \infty$ at $t=t_s$ \cite{DDGH}. These special models have been checked against data in Ref. \cite{laszlo}.

For both SFS and FSFS models described in terms of the scale factor (\ref{sf2}), the evolution begins with the standard big-bang
singularity at $t=0$, where $a=0$, and finishes at an exotic singularity for $t=t_s$, where $a=a_s\equiv a(t_s)$ is a constant. In terms of the rescaled time $y$ we have $a(1) = a_s$.

The standard Friedmann limit (i.e. models without an exotic singularity in future) of (\ref{sf2}) is achieved when $\delta \to 0$; hence $\delta$ is called
the ``non-standardicity" parameter. Additionally, notwithstanding Ref. \cite{barrow04} and in agreement with the field equations (\ref{rho})-({\ref{p}), $\delta$ can be
both positive and negative leading to an acceleration or a deceleration of the universe, respectively.

It is important to our discussion that the asymptotic behaviour of the scale factor (\ref{sf2}) close to a big-bang singularity at $t=0$ is given by a simple power-law $a_{\rm BB} = a_s y^m$, simulating the behavior of flat $k=0$ barotropic fluid models with $m = 2/[3(w+1)]$, where $w$ is a barotropic index.

Both SFS and FSFS scenarios consist of two components such as a nonrelativistic matter, and an exotic fluid (which we will further call dark energy with the energy density $\rho_{DE}$) which drives a singularity. We consider the case of the noninteracting components $\rho_m$ and $\rho_{DE}$  which both obey independently their continuity equations of the type (\ref{conser}). The evolution of both ingredients is independent. Nonrelativistic matter scales as $a^{-3}$, i.e.
\be
\rho_m=\Omega_{m0}\rho_0\left(\frac{a_0}{a}\right)^3~~,
\ee
and the evolution of the exotic (dark energy) fluid $\rho_{DE}$, can be determined by taking the difference between the total energy density $\rho$, which enters the  Friedmann equation (\ref{rho}), and the energy density of nonrelativistic matter, i.e.
\be
\rho_{DE}=\rho-\rho_m~~.
\ee
In fact, $\rho_{DE}$ component of the content of the Universe is responsible for an exotic singularity at $t\rightarrow t_s$. The dimensionless energy densities are defined in a standard way as
\be
\Omega_m=\frac{\rho_m}{\rho}~~, \hspace{0.3cm} \Omega_{DE}=\frac{\rho_{DE}}{\rho}~~,
\ee
and so for a dimensionless exotic dark energy density we have
\be
\label{OmDE}
\Omega_{DE}=1-\Omega_{m0}\frac{H_0^2}{H^2(t)}\left(\frac{a_0}{a(t)}\right)^3=1-\Omega_{m}.
\ee
We can then define the barotropic index of the equation of state for the dark energy as
\be
\label{wDE}
w_{DE}=p_{DE}/ \rho_{DE}~~,
\ee
and the effective barotropic index of the total equation of state is given by Eq. (\ref{weff}).

\section{Redshift drift}
\setcounter{equation}{0}
\label{drift}

We proceed within the framework of Friedmann cosmology, and consider an observer located at $r=0$ at coordinate time $t=t_0$. The observer receives a light ray emitted at $r=r_1$ at coordinate time
$t=t_1$ and, according to (\ref{sf2}), its redshift is given by
\be
\label{redshift}
1+z=\frac{a(t_0)}{a(t_1)} = \frac{\delta +
\left(1 - \delta \right) \left(\frac{t_0}{t_s}\right)^m - \delta \left( 1 - \frac{t_0}{t_s} \right)^n}
{\delta + \left(1 - \delta \right) \left(\frac{t_1}{t_s}\right)^m - \delta \left( 1 - \frac{t_1}{t_s}
\right)^n}~.
\ee
We then have a standard null geodesic equation
\be
\label{geod}
\int_0^{r_1} \frac{dr}{\sqrt{1-kr^2}} = \int_{t_1}^{t_0}
\frac{cdt}{a(t)}~,
\ee
with the scale factor $a(t)$ given by (\ref{sf2}). For a flat $k=0$ Friedmann model we can write down the radial coordinate of an observer in any of the forms below
\be
\label{radial}
r_1 = \int_{t_1}^{t_0} \frac{cdt}{a(t)} = \int_{a_1}^{a_0} \frac{cda}{H(a)a^2} = \int_{0}^{z} \frac{cdz}{H(z)a_0}~.
\ee
In (\ref{radial}) $a_0 \equiv a(t_0)$ and the transition from the integral of $da$ to the integral of $dz$ was given by the application of the definition of redshift (\ref{redshift}). Besides, due a lack of an analytic form for the equation of state for SFS models the function $H(z)$ can only be given by a formula which involves an integral over $z$, as follows
\be
\label{H(z)}
H^2(z) = H_0^2 \Omega_{e} (1+z)^3 \exp{\left[ \int_{0}^{z}
dz' \frac{2q(z') - 1}{1+z'} \right]}~,
\ee
in (\ref{radial}) with $\Omega_{e}$ being the density parameter of an exotic singularity driven dark energy \cite{AIP2010}. It is easy to notice that in the limit $\delta \to 0$ and $m=2/3$ one has $H(z) = H_0^2 \Omega_{m0} (1+z)^3$, as for dust matter dominated case. Let us recall that the standard formula for the models which also includes the dark energy component $\Omega_{w0}$ reads as \cite{DDGH}
\be
\label{standHz}
H^2(z) = H_0^2 \left[\Omega_{m0} (1+z)^3 + \Omega_{w0} (1+z)^{3(w+1)} \right]~.
\ee
However, in our further calculations we will not be expressing $H(z)$ in either forms (\ref{H(z)}) or (\ref{standHz}), using an explicit form of $a(t)$ as in (\ref{radial}) instead.

Now we consider the redshift drift effect in cosmology \cite{sandage}. In order to do that we assume that the source does not possess any peculiar velocity, so that it maintains a fixed comoving coordinate $dr=0$. The light emitted by the source at two different moments of time $t_e$ and $t_e+\delta t_e$ in VSL universe will be observed at $t_o$ and $t_o+\delta t_o$ related by
\be
\int_{t_e}^{t_o}\frac{c dt}{a(t)}=\int_{t_e+\Delta t_e}^{t_o+\Delta t_o}\frac{c dt}{a(t)}~,
\ee
which for small $\Delta t_e$ and $\Delta t_o$ transforms into
\be
\label{rel}
\frac{\Delta t_e}{a(t_e)}=\frac{\Delta t_o}{a(t_o)}~.
\ee
The redshift drift is defined as \cite{sandage}
\begin{eqnarray}
\label{redshiftdrift}
\Delta z = z_e - z_0 = \frac{a(t_0 + \Delta t_0)}{a(t_e + \Delta t_e)} - \frac{a(t_0)}{a(t_e)}~,
\end{eqnarray}
which can be expanded in series and to first order in $\Delta t$ (cf. Ref. \cite{DriftVSL}) as
\bea
&& \Delta z = \frac{a(t_0) + \dot{a}(t_0)\Delta t_0}{a(t_e) + \dot{a}(t_e)\Delta t_e} - \frac{a(t_0)}{a(t_e)} \nonumber \\
&& \approx \frac{a(t_0)}{a(t_e)} \left[ \frac{\dot{a}(t_0)}{a(t_0)} \Delta t_0 - \frac{\dot{a}(t_e)}{a(t_e)} \Delta t_e \right]~~.
\eea
Using (\ref{rel}) we have \cite{sandage}
\be
\label{exdrift}
\Delta z=\Delta t_0\left[ H_0(1+z) - H(t(z))\right]= (1+z)\frac{\Delta v}{c}~~,
\ee
where $\Delta v$ is the velocity shift and $H(t(z))$ is given by (\ref{H(z)}) with the application of (\ref{redshift}).

Several forthcoming facilities are expected to carry out measurements of the redshift drift, using independent techniques and probing different redshift ranges. In what follows our discussion will be based on the measurements carried out by ELT-HIRES, a high-resolution ultra-stable spectrograph for the E-ELT \cite{HIRES} which will carry out measurements using the Lyman-$\alpha$ forest \cite{loeb}. The behaviour of the spectroscopic velocity uncertainty is expected to take the following form \cite{E-ELT}
\be
\label{error}
\sigma_v=1.35\left(\frac{S/ N}{2370}\right)^{-1}\left(\frac{N_{qso}}{30}\right)^{-1/2}\left(\frac{1+z_{qso}}{5}\right)^{-1.7},
\ee
where $S/N$ is the signal to noise ratio for the spectra, $N_{qso}$ is the number of the absorbing systems and $z_{qso}$ are their redshifts. This formula is valid for $z\leq4$. We assumed constant ratio $S/N=3000$, 40 systems which are uniformly divided into 4 subsystems at $z=2,\ 3,\ 4,\ 5$ and we take the time between the observations $\Delta t_0 = 20$ years.

While these E-ELT measurements are particularly promising, since they span a large redshift lever arm that cannot otherwise be probed (and are orthogonal, in the relevant parameter space, to those obtained by conventional probes \cite{orthog}) we should also mention that redshift drift measurements at low redshift ($z<1$) are also expected from the SKA (Square Kilometre Array) \cite{SKA} and from intensity mapping experiments like CHIME (The Canadian Hydrogen Intensity Mapping Experiment) \cite{CHIME}. On a longer timescale, low-redshift measurements should also be carried out by the proposed gravitational wave interferometers DECIGO/BBO (DECi-hertz Interferometer Gravitational Wave Observatory/Big Bang Observer) \cite{DECIGO}.

\section{Results}
\setcounter{equation}{0}
\label{results}
\begin{center}
  \begin{table}
   \begin{tabular}{lcccc}
     & {\bf $m$}     &    {\bf $\delta$ } & {\bf $n$}& {\bf $t_0/t_s$}         \\\hline
     SFS 1 &$2/3$  &  $-0.43$ & $1.9999$ & $0.99$    \\
 SFS 2 &$2/3$  &  $0.0$   & $1.9999$ & $0.99$ \\
 SFS 3 &$0.749$ &  $-0.45$  & $1.99$ & $0.77$  \\
 FSFS 1&$0.56$    &  $0.42$ & $0.8$  & $0.96$  \\
 FSFS 2&$2/3$     &  $0.0$  & $0.7$  & $0.79$  \\
 FSFS 3&$2/3$     &  $0.24$ & $0.7$  & $0.96$  \\
 FSFS 4&$1.15$    &  $7.5$  & $0.81$ & $0.51$  \\
  \end{tabular}
  \caption{The values of the parameters for the models which are investigated in this paper. As for the standard cosmological parameters we take $[H_0]=67.3$ $km/s/Mpc$ and $\Omega_{m0} = 0.315$ as supported by the Planck data \cite{planck13}.}
  \label{tabelka}
 \end{table}
\end{center}

We applied the formula (\ref{exdrift}) to investigate the effect of redshift drift for selected SFS and FSFS models with the parameter values given in Table \ref{tabelka}. In Figs. \ref{SFSdeltaz}-\ref{sfswde} we present the plots of the redshift drift (\ref{exdrift}), the density parameter (\ref{OmDE}), the Hubble function (\ref{H(z)}), and the dark energy barotropic index (\ref{wDE}) for three different SFS models as well as for the $\Lambda$CDM model. The model SFS1 has the same set of parameters as in Ref. \cite{DHD}, SFS2 is just the standard dust (Einstein-de Sitter) limit $\delta \to 0$ of the SFS models, and SFS3 is plotted for the observationally best-fit region (supernovae, BAO, shift parameter) of the parameters as given in Refs. \cite{GHDD,DDGH} and gives a complementary test of SFS models. As can be seen from the plots, SFS3 can mimic $\Lambda$CDM model, while SFS1 behaves differently than $\Lambda$CDM for large redshifts. SFS1 has a characteristic peak just close to $z \sim 0$ which gives positive values of the drift (up to z $\approx 1$) and further falls down sharply to negative values of $\Delta z$ then behaving in a similar way as the dust Friedmann model (here marked as SFS2). Thus such models can be tested observationally to exclude one of them. In fact, for ELT-HIRES only the region $z > 1.7$ is reachable, but smaller redshifts might be reachable by future radiotelescopes
\cite{SKA,CHIME} as well as space-borne gravitational telescopes \cite{DECIGO}. Figure \ref{sfswde} nicely shows that SFS1 and SFS3 models mimic dark energy for smaller redshifts $(z<2)$, while they behave like dust in the limit of large redshifts. For SFS2, one has $w_{DE}=0$.

\begin{figure}
\begin{center}

\includegraphics[width=9.cm]{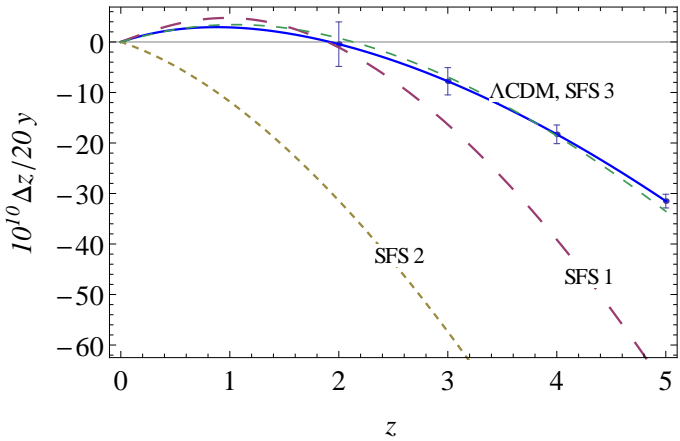}

\caption{The plot of redshift drift (\ref{exdrift}) for sudden future singularity models. Three models are presented: SFS1 with $m=2/3, n=1.9999, \delta=-0.43, (t_0/t_s) = 0.99$; SFS2 (standard dust Friedmann) with $m=2/3, n=1.9999, \delta=0.0, (t_0/t_s)=0.99$; and SFS3 with $m=0.749, n=1.99, \delta=-0.45, (t_0/t_s)=0.77$. The uncertainties given by the formula (\ref{error}) are taken from Ref. \cite{E-ELT}.}
\label{SFSdeltaz}
\end{center}
\end{figure}

\begin{figure}
\begin{center}

\includegraphics[width=8.9cm]{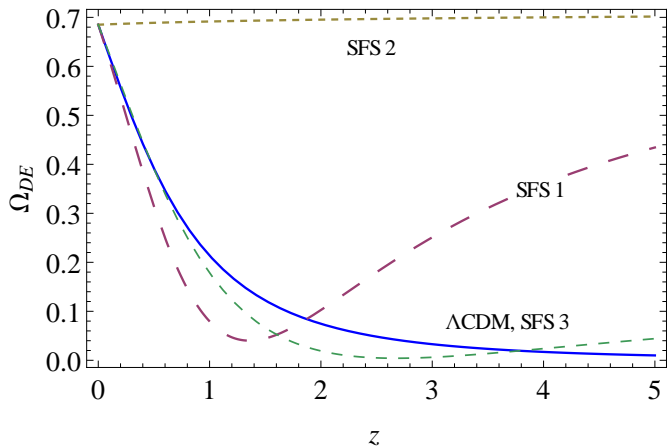}

\caption{The plot of the density parameter (\ref{OmDE}) for the sudden future singularity models with the same parameters as in Fig. \ref{SFSdeltaz} (cf. Table \ref{tabelka}).}
\label{sfsomde}
\end{center}
\end{figure}

\begin{figure}
\begin{center}

\includegraphics[width=8.9cm]{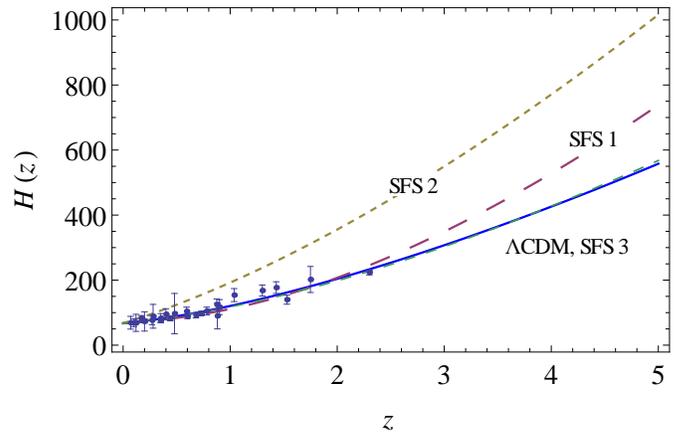}

\caption{The plot of the Hubble function (\ref{H(z)}) for the sudden future singularity models with the same parameters as in Fig. \ref{SFSdeltaz} (cf. Table \ref{tabelka}). The data points are taken from Ref. \cite{farooq}.}
\label{sfsHz}
\end{center}
\end{figure}

\begin{figure}
\begin{center}

\includegraphics[width=8.9cm]{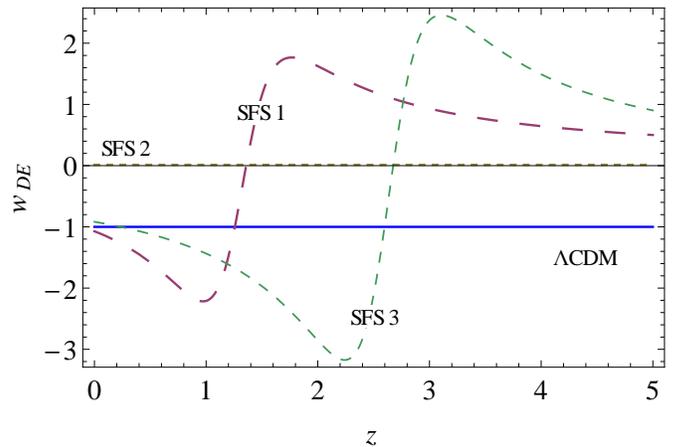}

\caption{The plot of the dark energy barotropic index (\ref{wDE}) for the sudden future singularity models with the same parameters as in Fig. \ref{SFSdeltaz}.}
\label{sfswde}
\end{center}
\end{figure}

In Figs. \ref{FSFSdeltaz}-\ref{fsfswde} we present the plots of the redshift drift (\ref{H(z)}), the density parameter (\ref{OmDE}), the Hubble function (\ref{H(z)}), and the dark energy barotropic index (\ref{wDE}) of the four different FSFS models as well as for the $\Lambda$CDM model. The model FSFS1 has similar set of parameters as in Ref. \cite{JCAP12}, FSFS2 is just the standard dust (Einstein-de Sitter) limit $\delta \to 0$ of the SFS models, while FSFS3 and FSFS4 are plotted for some other observationally allowed values of the parameters. We note that model FSFS4 provides a reasonable fit with other datasets \cite{JCAP12}, but has an $H(z)$ behavior that is very different from the observed data as shown in the Fig. \ref{fsfsHz}, and can be considered as a more extreme case of the dust-filled model FSFS2. This highlights the usefulness of combining different observables. The plots show that FSFS1 can mimic $\Lambda$CDM model, FSFS3 is relatively close to $\Lambda$CDM, while FSFS4 differs from $\Lambda$CDM significantly. In particular, FSFS4 has a very sharp and a narrow peak near to $z \sim 0$, and is very different from $\Lambda$CDM. This sharp peak differs FSFS4 also from the dust-filled Friedmann model (FSFS2). All this allows to differentiate FSFS3 and FSFS4 models from $\Lambda$CDM models.
\begin{figure}
\begin{center}

\includegraphics[width=8.9cm]{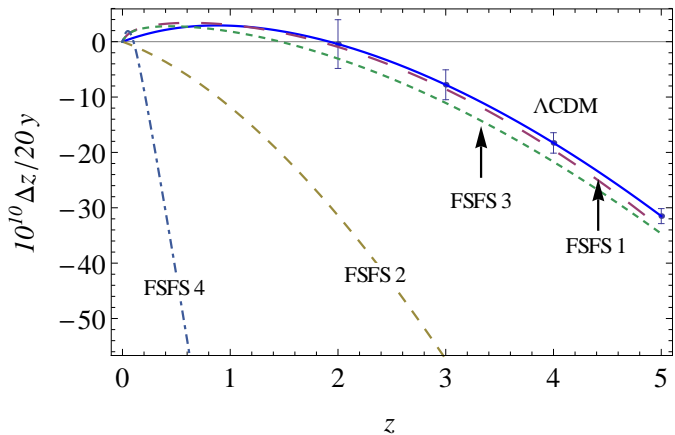}
\caption{The plot of redshift drift (\ref{exdrift}) for finite scale factor singularity models. Four models are presented: FSFS1 with $m=0.56, n=0.8, \delta=0.42, (t_0/t_s)=0.96$; FSFS2 (dust Friedmann) with $m=2/3, n=0.7, \delta=0.0, (t_0/t_s)=0.79$; FSFS3 with $m=2/3, n=0.7, \delta=0.24, (t_0/t_s)=0.96$, and FSFS4 with $m=1.15, n=0.81, \delta=7.5,y_0=0.51$. The uncertainties given by formula (\ref{error}) are taken from Ref. \cite{E-ELT}.}
\label{FSFSdeltaz}
\end{center}
\end{figure}

\begin{figure}
\begin{center}

\includegraphics[width=8.9cm]{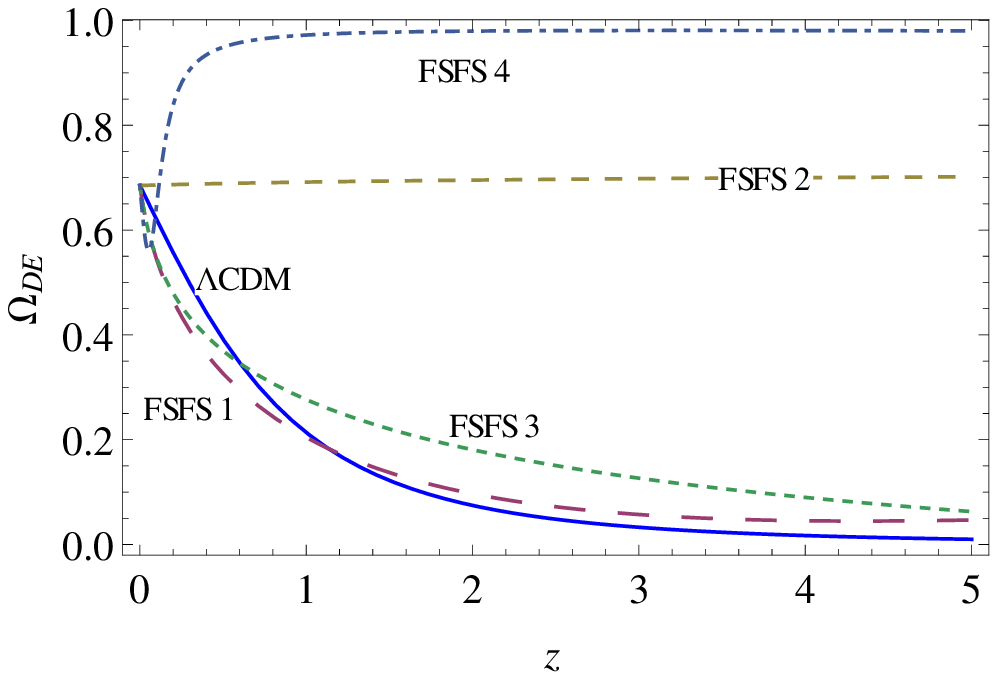}

\caption{The plot of the density parameter (\ref{OmDE}) for the finite scale factor singularity models with the same parameters as in Fig. \ref{FSFSdeltaz} (cf. Table \ref{tabelka}).}
\label{fsfsomde}
\end{center}
\end{figure}
\begin{figure}
\begin{center}

\includegraphics[width=8.9cm]{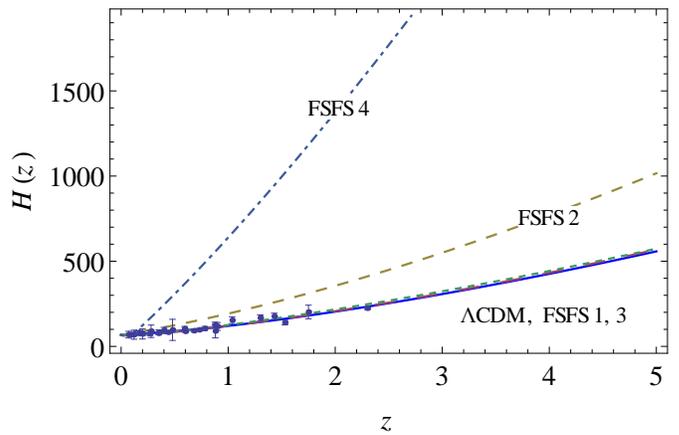}

\caption{The plot of the Hubble function (\ref{H(z)}) for the finite scale factor singularity models with the same parameters as in Fig. \ref{FSFSdeltaz} (cf. Table \ref{tabelka}). The data points are taken from Ref. \cite{farooq}.}
\label{fsfsHz}
\end{center}
\end{figure}
\begin{figure}
\begin{center}

\includegraphics[width=8.9cm]{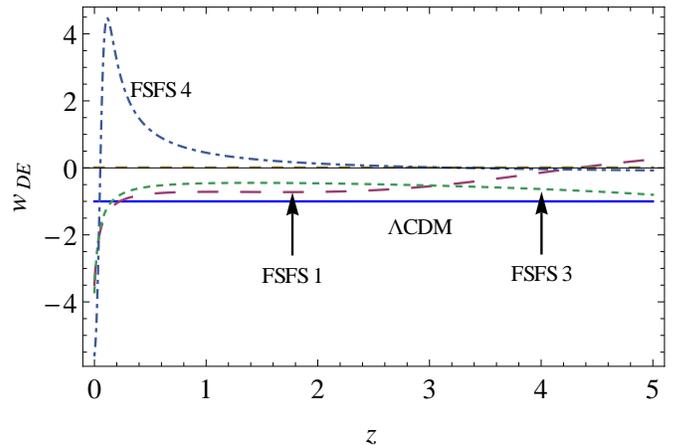}

\caption{The plot of the barotropic dark energy parameter (\ref{wDE}) for the finite scale factor singularity models with the same parameters as in Fig. \ref{FSFSdeltaz} (cf. Table \ref{tabelka}).}
\label{fsfswde}
\end{center}
\end{figure}

\section{Conclusions}
\label{rac}
\setcounter{equation}{0}

We have presented some viable examples of exotic singularity universes and studied the redshift drift effect which allows to look at two different past light cones of an observer placed on Earth separated by the time of 20 years. Two particular models have been investigated: the ones which admit sudden future (or pressure) singularities and another ones which admit finite scale factor (of energy density and pressure) singularities.

Our analysis shows that the range of redshift signal is made discrepant from $\Lambda$CDM is model-dependent: some models in this class are easier to distinguish at low redshifts (within reach od SKA or CHIME) while for others to high redshifts probed by ELT-HIRES will be ideal.

We have shown that both these models, for special values of the parameters (chosen for models marked as SFS3 and FSFS1), can mimic the redshift drift behavior of $\Lambda$CDM models. In other words, the matter which is responsible for the singularities behaves as dark energy. As for other values of the parameters (models SFS1, FSFS3) the redshift drift test can play the role of a differential test between these models and $\Lambda$CDM, though leaving them still good candidates for dark energy.

\section{Acknowledgements}

\indent The research of T.D. was financed from the state budget for education grant No 0218/IP3/2013/72 (years 2013-2015) and of M.P.D. by the National Science Center Grant DEC-2012/06/A/ST2/00395. C.J.M. and P.E.V. are supported by project PTDC/FIS/111725/2009 from FCT, Portugal. C.J.M. is supported by an FCT Research Professorship, contract reference IF/00064/2012, funded by FCT/MCTES (Portugal) and POPH/FSE (EC).\\
\indent Part of the results reported in this work were performed using the HPC cluster HAL9000 of the Computing Centre of the Faculty of Mathematics and Physics at the University of Szczecin.

\end{document}